\documentclass[aip,apl, reprint, twocolumn,amsmath,amssymb,superscriptaddress]{revtex4-1}

\usepackage{graphicx}
\usepackage{dcolumn}
\usepackage[ansinew]{inputenc}
\usepackage{bm}
\usepackage{mathrsfs}
\usepackage{amsfonts}
\usepackage{amssymb}
\usepackage{hyperref}

\newcommand{\bra}[1]{\langle #1|}
\newcommand{\ket}[1]{|#1\rangle}

\begin{document}

\title{Imaging of microwave fields using ultracold atoms}

\author{Pascal Böhi}
\author{Max F. Riedel}
\author{Theodor W. Hänsch}
\affiliation{Max-Planck-Institut für Quantenoptik und Fakultät für Physik der Ludwig-Maximilians-Universität, 80799 München, Germany}
\author{Philipp Treutlein}\email[E-mail: ]{philipp.treutlein@unibas.ch}
\affiliation{Max-Planck-Institut für Quantenoptik und Fakultät für Physik der Ludwig-Maximilians-Universität, 80799 München, Germany}
\affiliation{Departement Physik der Universität Basel, CH-4056 Basel, Switzerland}


\begin{abstract}
We report a technique that uses clouds of ultracold atoms as sensitive, tunable, and non-invasive probes for microwave field imaging with micrometer spatial resolution. The microwave magnetic field components drive Rabi oscillations on atomic hyperfine transitions whose frequency can be tuned with a static magnetic field. Readout is accomplished using state-selective absorption imaging. 
Quantitative data extraction is simple and it is possible to reconstruct the distribution of microwave magnetic field amplitudes and phases. 
While we demonstrate 2d imaging, an extension to 3d imaging is straightforward. We use the method to determine the microwave near-field distribution around a coplanar waveguide 
integrated on an atom chip.

\end{abstract}
\maketitle

Today, Monolithic Microwave Integrated Circuits (MMICs) are of great importance in science and technology. In particular, they constitute key building blocks of today's communication technology.\cite{Robertson01} MMICs also serve as main components of superconducting quantum processors.\cite{DiCarlo09}
In our group, a simple MMIC structure has recently been used as a tool for quantum coherent manipulation of ultracold atoms on an atom chip.\cite{Boehi09a}

Function and failure analysis is of crucial importance for the design of MMICs as well as for simulation verification.\cite{Boehm94} External port measurements (e.g.\ using a network analyzer) offer only limited insight. The microwave (mw) near-field distribution on the device gives much more information, enabling specific improvement. Therefore, different methods have been developed to measure the spatial distribution of mw near-fields.\cite{imagingpapers} 
These methods use diverse physical effects to measure the mw electric or magnetic field. They have in common that they scan the field distribution point-by-point. 

Here we propose and experimentally demonstrate a highly parallel method that allows for non-invasive and complete (amplitudes and phases) imaging of the mw magnetic field distribution  using clouds of ultracold atoms.\cite{NatureInsight02} In this method, the mw magnetic field drives resonant Rabi oscillations\cite{Gentile89} between two atomic hyperfine levels that can be detected using state-selective absorption imaging.\cite{Matthews98} The method offers $\mu\mathrm{m}$ spatial resolution and a mw magnetic field sensitivity in the $10^{-8}~\mathrm{T}$ range at frequencies of a few GHz. It is a frequency-domain, single-shot technique to measure a 2d field distribution. The method can be extended to measure 3d distributions slice by slice. Data extraction is simple, it offers a high dynamic range, and it is intrinsically calibrated since only well-known atomic properties enter in the analysis.

\begin{figure}[htbp]
	\centering
		\includegraphics[width=0.45\textwidth]{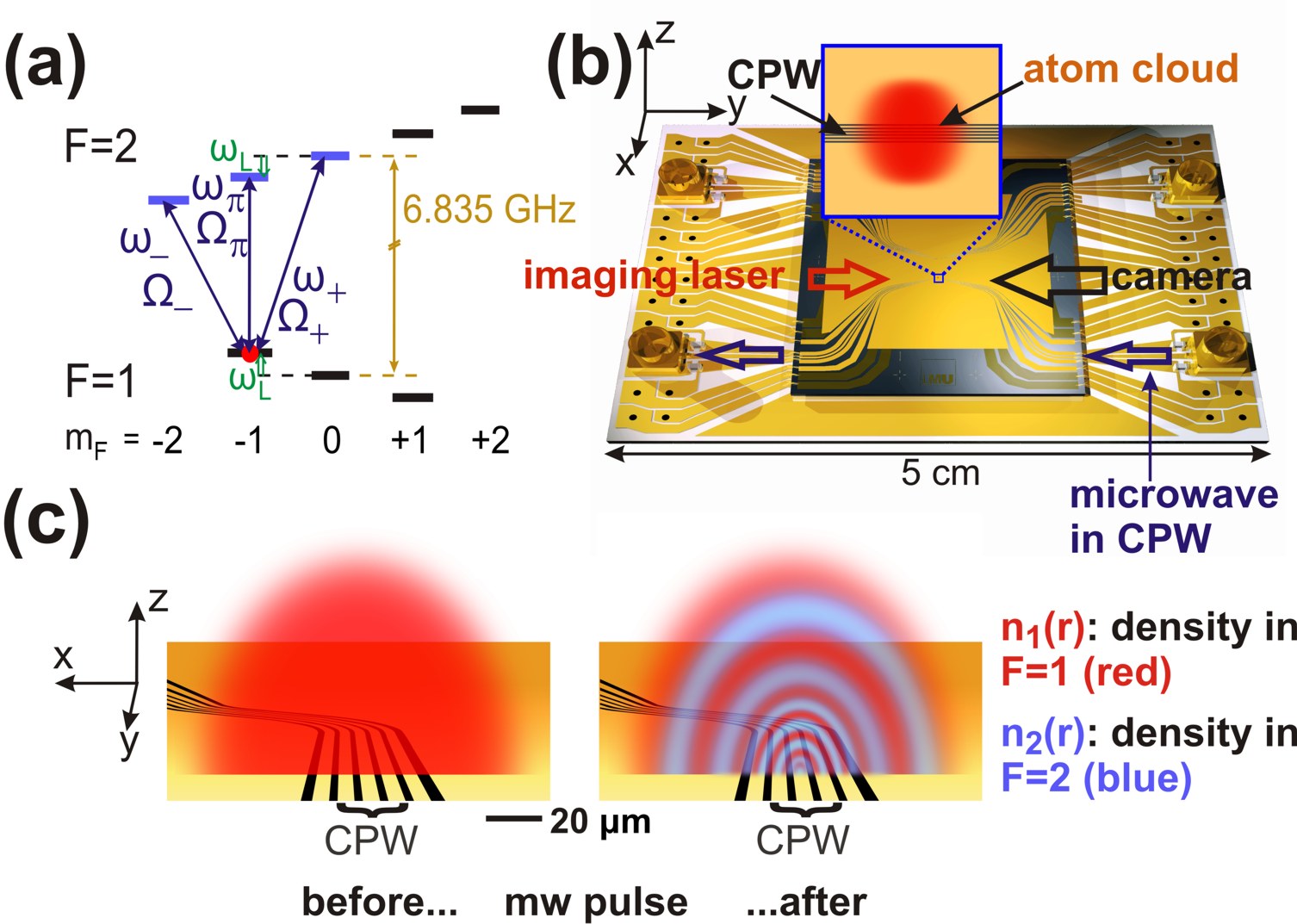}
	\caption{Schematic of the experiment. (a) Ground state hyperfine levels of $^{87}\mathrm{Rb}$ atoms in a static magnetic field. Initially, the atoms are trapped in state $\ket{F, m_{F}}=\ket{1,-1}$. The three relevant transitions $\ket{1,-1} \rightarrow\ket{2,m_{2}}$, ($m_{2}=-2,-1,0$) are indicated. The corresponding transition frequencies $\omega_\gamma$, ($\gamma=-, \pi, +$) are split by $\omega_{\mathrm{L}}$ due to the Zeeman effect. 
 The Rabi frequencies $\Omega_\gamma$ are indicated. 
(b) 
The atom chip. Inset: Atom cloud near the coplanar waveguide (CPW) structure whose mw magnetic field is examined. The three inner wires constitute the CPW.  
(c) Experimental sequence. Left: The trap is switched off and the atom cloud expands. 
Right: A mw pulse is applied to the CPW, resonant with one of the transitions $\omega_{\gamma}$. Its magnetic field of amplitude $\mathbf{B}(\mathbf{r})$ drives Rabi oscillations with position-dependent $\Omega_\gamma(\mathbf{r})$ between $\ket{1,-1}$ (red) and the corresponding state $\ket{2,m_{2}}$ (blue). The resulting atomic density distribution $n_{1}(\mathbf{r})$ ($n_{2}(\mathbf{r})$) in $F=1$ ($F=2$) 
is detected. 
From $n_i(\mathbf{r})$, we reconstruct $\Omega_\gamma(\mathbf{r})$. Several such measurements on the three transitions are combined to reconstruct $\mathbf{B}(\mathbf{r})$.}
	\label{fig:Fig1}
\end{figure}

For the proof-of-principle experiment presented here, we use our atom chip setup,\cite{Boehi09a} see Fig.~\ref{fig:Fig1}. The chip has integrated mw coplanar waveguide structures that were designed for quantum manipulation of ultracold atoms.\cite{Boehi09a} 
To demonstrate our method, we analyze the mw magnetic field near this structure. In general, the device to be tested does not have to be integrated on an atom chip.

Our method works as follows. On the room-temperature atom chip, we prepare clouds of $^{87}\mathrm{Rb}$ atoms at a temperature of $T \simeq 5~\mathrm{\mu K}$ using laser and evaporative cooling techniques.\cite{Boehi09a,NatureInsight02} 
The magnetically trapped atoms are initially in the ground state hyperfine sublevel $\ket{F,m_F}=\ket{1,-1}$, see Fig.~\ref{fig:Fig1}a. 
The trap is moved close to the mw structure to be characterized, where it is switched off and the atoms are released to free fall. 
During a hold-off time $dt_{\mathrm{ho}}$, the cloud drops due to gravity and expands due to its thermal velocity spread, filling the region to be imaged (Fig.~1b+c).
We maintain a homogeneous static magnetic field $\mathbf{B}_{0}$ of order $10^{-4}$~T. It provides the quantization axis and splits the frequencies $\omega_\gamma$, ($\gamma=-,\pi,+$) of the three hyperfine transitions $\ket{1,-1} \rightarrow \ket{2,m_{2}}$, ($m_{2}=-2,-1,0$) by $\omega_{\mathrm{L}}=\mu_B B_{0}/ 2 \hbar$ due to the Zeeman effect (cf.\ Fig.~1a). 
A mw signal on the MMIC is subsequently switched on for a duration $dt_{\mathrm{mw}}$ (typically some tens of $\mu$s). 
We select one of the transitions by setting the mw frequency $\omega=\omega_\gamma$.
The mw magnetic field $\mathscr{B}(\mathbf{r},t) = \tfrac{1}{2} \left[ \mathbf{B}(\mathbf{r})e^{-i\omega t} + \mathbf{B}^*(\mathbf{r})e^{i\omega t}\right]$  couples to the atomic magnetic moment and drives Rabi oscillations of frequency $\Omega_\gamma(\mathbf{r})$ on the resonant transition.
For the three transitions of interest, 
\begin{eqnarray}
 \Omega_{-}(\mathbf{r})  &=& -\sqrt{3}\frac{ \mu_{B}}{\hbar}B_{-}(\mathbf{r})e^{-i \phi_{-}(\mathbf{r})}, \label{eq:OmegaSigMin}\\
 \Omega_{\pi}(\mathbf{r})    &=& -\sqrt{\frac{3}{4}}\frac{ \mu_{B}}{\hbar}B_{\pi}(\mathbf{r})e^{-i \phi_{\pi}(\mathbf{r})}, \label{eq:OmegaPi} \\
 \Omega_{+}(\mathbf{r}) &=& \sqrt{\frac{1}{2}}\frac{ \mu_{B}}{\hbar}B_{+}(\mathbf{r})e^{-i \phi_{+}(\mathbf{r})} . \label{eq:OmegaSigPlus}
\end{eqnarray}
Here, $B_{\pi}$ and $\phi_\pi$ are the real-valued amplitude and phase of the component of $\mathbf{B}$ parallel to $\mathbf{B}_{0}$, and $B_{+}, \phi_+$ ($B_{-}, \phi_-$) are the corresponding quantities for the right (left) handed circular polarization component in the plane perpendicular to $\mathbf{B}_{0}$, see Supplementary Information.
After the mw pulse, a spatial pattern of atomic populations in $F=1$ and $F=2$ results, see Fig.~\ref{fig:Fig1}c. The probability to detect an atom in $F=2$ is
\begin{equation}
p_{2}(\mathbf{r})\equiv \frac{n_{2}(\mathbf{r})}{n_{1}(\mathbf{r})+n_{2}(\mathbf{r})}=\sin^{2}\left[ \tfrac{1}{2} |\Omega_{\gamma}(\mathbf{r})| dt_{\mathrm{mw}} \right]  \label{eq:p2}.
\end{equation} 
Here, $n_{1}(\mathbf{r})$ ($n_{2}(\mathbf{r})$) is the density of atoms in $F=1$ ($F=2$), which can be measured using state-selective absorption imaging.\cite{Matthews98}  
From $p_2(\mathbf{r})$ we can reconstruct $|\Omega_{\gamma}(\mathbf{r})|$ and thus the spatial distribution of the resonant mw polarization component $B_\gamma (\mathbf{r})$, as shown below.
\begin{figure}[tb]\centering
		\includegraphics[width=0.45\textwidth]{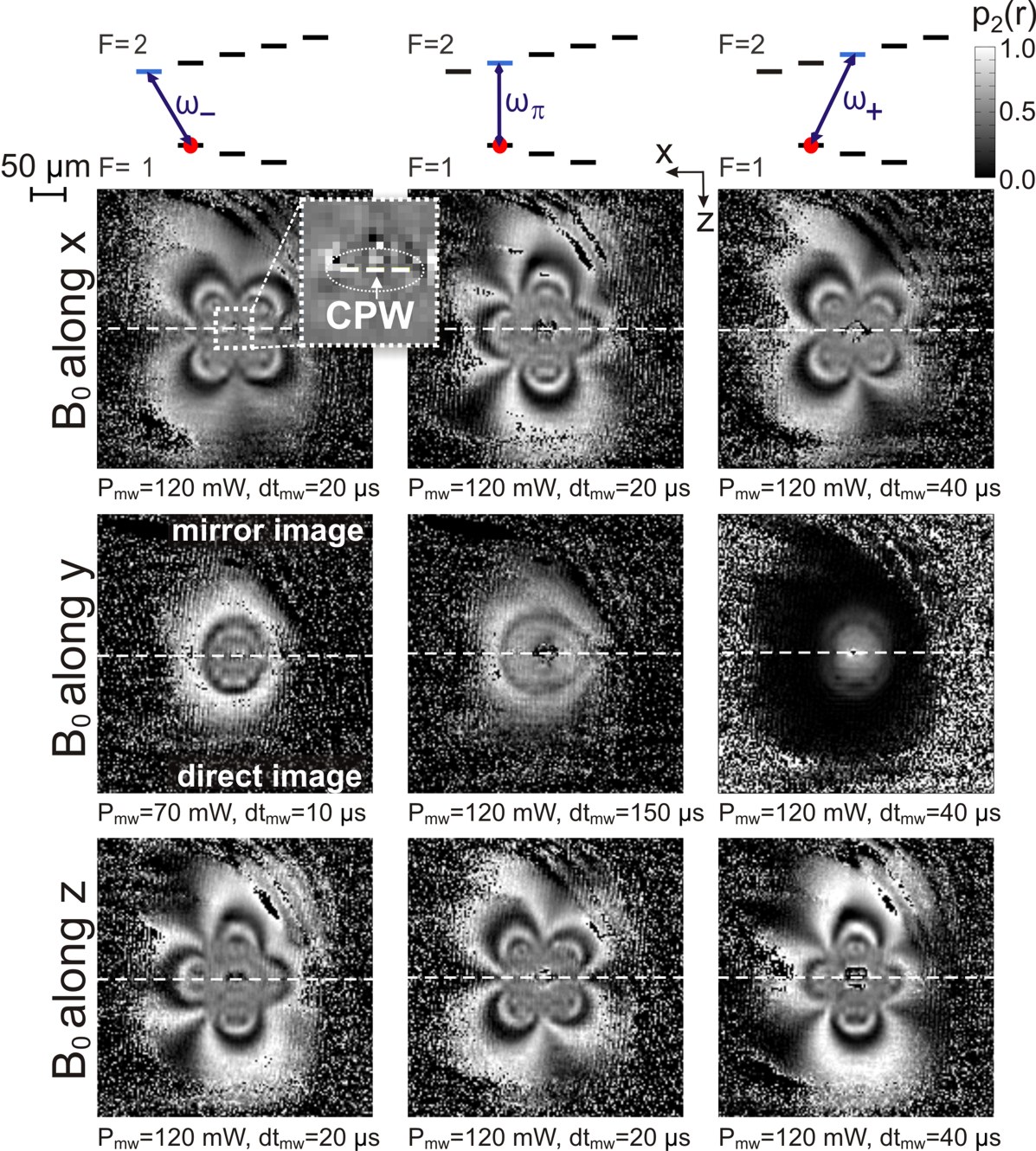}
	\caption{Imaging of mw magnetic field components near the CPW of Fig.~\ref{fig:Fig1}. The images show the measured probability $p_{2}(\mathbf{r})$ to find an atom in $F=2$ after applying the mw pulse. Columns correspond to measurements on the three different transitions $\omega_\gamma$, rows to three different orientations of $\mathbf{B}_{0}$.  
The imaging beam is reflected from the chip surface at an angle of $2^\circ$. As a result, on each picture, a direct image and its reflection on the surface are visible. The dashed line separates the two. 
No atoms are visible in the very center of the image because the CPW structures distort the imaging beam.
The mw power launched into the CPW, $P_{\mathrm{mw}}$, and the mw pulse duration $dt_{\mathrm{mw}}$ are indicated.  $dt_\mathrm{ho}$ varies between $1$ -- $2$~ms.
The noise in the image periphery corresponds to regions without atoms. Images are averaged over several experimental runs (15 to 130).
}
	\label{fig:Fig2}
\end{figure}

\begin{figure}[tb] \centering
		\includegraphics[width=0.45\textwidth]{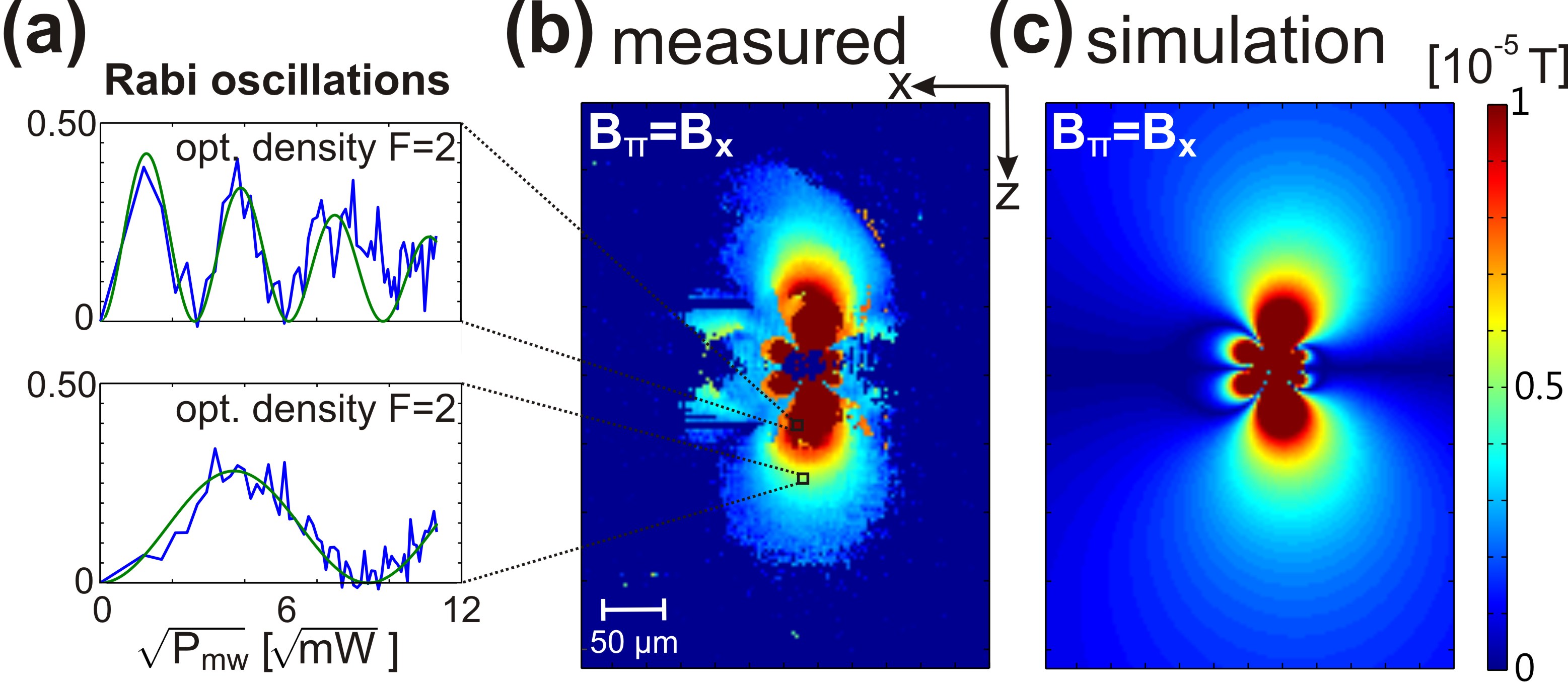}
	\caption{Measured mw magnetic field component $B_x(\mathbf{r})$ near the CPW and comparison with a simulation. The measurement is performed with $\omega=\omega_\pi$ and $\mathbf{B}_0$ along $x$, and the \textit{movie method} (see text) is used to extract $B_\pi(\mathbf{r}) \equiv B_x(\mathbf{r})$.
	(a) Rabi oscillations at two exemplary pixels of the image, recorded by varying $P_\mathrm{mw}$ at fixed $dt_\mathrm{mw}=20~\mu$s. The sinusoidal fits used to determine $|\Omega_\pi |$ and thus $B_\pi \equiv B_x$ as a function of $\sqrt{P_\mathrm{mw}}$ are shown. The observed decay of the oscillations is due to mw field gradients across the pixel.
	(b) Image of $B_x(\mathbf{r})$ at $P_{\mathrm{mw}}=120$~mW as obtained from the data.
	(c) Corresponding quasi-static simulation of $ B_x(\mathbf{r})$.\cite{Boehi09a} We find best agreement with (b) if we allow for a 10\% asymmetry between the currents on the two CPW ground wires and assume induced currents in the two wires next to the CPW grounds with an amplitude of 2\% of the signal conductor current.
	}
	\label{fig:Fig3}
\end{figure}

For a reconstruction of $\mathbf{B}(\mathbf{r})$, we measure $B_{\pi}(\mathbf{r})$, $B_{+}(\mathbf{r})$, and $B_{-}(\mathbf{r})$ with $\mathbf{B}_{0}$ pointing along the $x$, $y$, and $z$ axis. In each of the nine measurements (cf.~Fig.~\ref{fig:Fig2}), we tune $\omega$ into resonance with the desired transition $\omega_{\gamma}$. Alternatively, to test a device at a given frequency $\omega$, one can achieve resonance by Zeeman-tuning of $\omega_\gamma$ via $B_{0}$.
The measurements allow a reconstruction of the amplitudes of all three cartesian components of $\mathbf{B}$ as well as their relative phases. The measurements of $B_\pi$ for the three orientations of $\mathbf{B}_{0}$ directly yield the amplitudes $B_x$, $B_y$, and $B_z$. The relative phases can be reconstructed from the other measurements, see Supplementary Information. It is also possible to measure the spatial dependence of the global phase of $\mathbf{B}$ using an interferometric method, see Supplementary Information. 

Absorption imaging integrates 
over the direction of propagation of the imaging laser beam. Structures to be characterized should therefore have a characteristic length scale along the beam larger than the size of the atom cloud. 
The cloud size can be adjusted through the magnetic trap frequencies, $T$, and $dt_{\mathrm{ho}}$. 

We experimentally demonstrate our method by measuring the mw magnetic field distribution near the coplanar waveguide (CPW) on our atom chip. 
The measurement is performed at a position where the CPW is translationally invariant along the imaging beam.
An overview of the data is shown in Fig.~\ref{fig:Fig2}. What appears in the images are essentially isopotential lines of the mw magnetic field components. Qualitative conclusions can be drawn directly by looking at the images -- e.g.\ the left/right asymmetry visible in Fig.~\ref{fig:Fig2} reveals that there is an asymmetry in the mw currents on the CPW wires. This confirms independent findings in our previous paper.\cite{Boehi09a}

It is possible to automatically extract $B_\gamma(\mathbf{r})$ from a single image. From $p_{2}(\mathbf{r})$ we can calculate $|\Omega_{\gamma}(\mathbf{r})|$ using Eq.~(\ref{eq:p2})
up to an offset of $n\, \pi/dt_\mathrm{mw}$, where $n$ is an integer. The offset for each point can be calculated by a \textit{ray-tracing method}, where rays are sent from the image periphery (where $|\Omega_\gamma| \, dt_\mathrm{mw} \ll 1$) through the desired point to the center of the mw structure (where $|\Omega_\gamma|$ is maximal). 
$n$ is given by the sum of the number of minima and maxima of $p_{2}(\mathbf{r})$ encountered on the ray.

Alternatively, we take series of $k$ image frames, scanning either the mw power $P_{\mathrm{mw}}$ or $dt_{\mathrm{mw}}$ (the \textit{movie method}). 
$k$ depends on the desired dynamic range, but can be as low as 10.  The time to record one frame is $14$~s, but could be reduced to $\leq3$~s.\cite{Farkas09}
For each image pixel, we thus obtain a sequence of $k$ datapoints showing Rabi oscillations, see Fig.~\ref{fig:Fig3}a. We fit a function $\propto \sin^{2}\left[ \tfrac{1}{2} |\Omega_\gamma (P_\mathrm{mw})|  dt_\mathrm{mw} \right]$ to the data, where $|\Omega_\gamma ( P_\mathrm{mw})| = a \sqrt{P_\mathrm{mw}}$ and $a$ is the fit parameter. From the fit, we determine $|\Omega_\gamma|$ and thus, via Eqs.~(\ref{eq:OmegaSigMin}) - (\ref{eq:OmegaSigPlus}),  $B_\gamma$ at this pixel for a given $P_\mathrm{mw}$. 
As an example, Fig.~\ref{fig:Fig3}b shows an image of the cartesian mw field component $B_x(\mathbf{r})$ near our CPW reconstructed in this way. Fig.~\ref{fig:Fig3}c shows a corresponding simulation. Comparing data and simulation, we obtain information about the current distribution on the CPW. 
The single-shot mw field sensitivity of our method is mainly determined by the interaction time $dt_{\mathrm{mw}}$. There is a trade-off between sensitivity and effective spatial resolution $s_{\mathrm{eff}}$. Longer $dt_{\mathrm{mw}}$ yields higher sensitivity, but at the same time the image blurs due to the movement of the atoms. For our parameters (optical resolution $s_{\mathrm{opt}}=4~\mathrm{\mu m}$, $dt_{\mathrm{mw}}=80~\mathrm{\mu s}$, $T=5~\mathrm{\mu K}$) and $dt_{\mathrm{ho}}=0$, we obtain $s_{\mathrm{eff}}=8.2~\mathrm{\mu m}$ and a mw magnetic field sensitivity of $2\times 10^{-8}$~T (Supplementary Information).

In a variant of the presented method, trapped atoms could be used as a scanning mw field sensor. For trapped atoms, $dt_{\mathrm{mw}}$ can be much longer without increasing $s_{\mathrm{eff}}$, thereby improving the field sensitivity. To record an image, the trap position has to be scanned from shot to shot of the experiment (Supplementary Information).


Our method is frequency selective. Individual components of a multi-tone signal could be resolved. The transition frequencies $\omega_{\gamma}$ can be adjusted via $B_{0}$. For $B_{0}$ up to $0.5$~T, which is a realistic effort e.g.\ for prototype testing, transition frequencies of $2.5$ - $14$~GHz are accessible with $^{87}\mathrm{Rb}$ (Supplementary Information). Note that for $B_{0}>0.1$~T we start entering the Paschen-Back regime, where the matrix elements of Eqs.~(\ref{eq:OmegaSigMin}) - (\ref{eq:OmegaSigPlus}) change and the theory has to be modified. 
Using other atomic species, different frequency ranges are accessible. 
Alternatively, a two-photon transition could be used for imaging, where two mw fields or a mw and a radio frequency field are applied to the atoms. The first field (frequency $\omega_{1}$) is applied externally with known spatial distribution, while the other field (frequency $\omega_{2}$) is to be imaged. Resonant Rabi oscillations occur for $\omega_{1}+\omega_{2}=\omega_{\gamma}$.
It is also possible to image off-resonant mw fields and other differential potentials using a Ramsey interferometry technique, see \cite{Ramsey56} and Supplementary Information.

Our technique can be extended to measure 3d distributions of $\mathbf{B}(\mathbf{r})$ slice by slice, either by using a gradient of $B_{0}$ such that only a slice of atoms is resonant with $\omega$, or by using a light sheet detection technique,\cite{Buecker09} where slices perpendicular to the camera line of sight are imaged. 
Using similar techniques, it is also possible to shape the atomic cloud to prepare a thin sheet of atoms. 

The method presented here seems promising for applications like prototype characterization. It allows for highly sensitive, parallel, high-resolution, and non-invasive imaging of the complete mw magnetic field distribution around an MMIC. 
Compact and portable systems for the preparation of ultracold atoms have been built,\cite{Farkas09}  and key components of such systems are commercially available. 

%

\onecolumngrid
\newpage
\section*{Imaging of microwave fields using ultracold atoms: Supplementary information}

\subsection{Rabi Frequencies}
We derive the Rabi frequencies \cite{Gentile89} for the resonant coupling of ground state hyperfine levels of $^{87}$Rb with a microwave (mw) field.
In the following, we consider an atom in a weak static magnetic field $\mathbf{B}_{0}$, so that the Zeeman splitting $\omega_L=\mu_B B_0/2\hbar$ is small compared to the zero-field splitting of $\omega_0 \simeq 2\pi \times 6.8$~GHz between the two hyperfine states $F=1$ and $F=2$ of the $5^{2}S_{1/2}$ electronic ground state of $^{87}$Rb.
The atom is initially prepared in the hyperfine sublevel $\ket{F,m_{F}}=\ket{1,-1}$, and the microwave frequency $\omega$ is resonant with one of the transitions $\ket{1,-1}\leftrightarrow \ket{2,m_2}$, ($m_2 = -2, -1, 0$) connecting to this level (see Fig.~1 of the main paper).

The real-valued mw magnetic field at position $\mathbf{r}=(x,y,z)$ in the fixed cartesian laboratory coordinate system is $\mathscr{B}(\mathbf{r},t) = \tfrac{1}{2} \left[ \mathbf{B}(\mathbf{r})e^{-i\omega t} + \mathbf{B}^*(\mathbf{r})e^{i\omega t}\right]$ with the complex phasor 
\[
\mathbf{B}(\mathbf{r})\equiv\left( 
\begin{array}{l}
	B_{x}(\mathbf{r})e^{-i\phi_{x}(\mathbf{r})} \\
	B_{y}(\mathbf{r})e^{-i\phi_{y}(\mathbf{r})} \\
	B_{z}(\mathbf{r})e^{-i\phi_{z}(\mathbf{r})} 
\end{array}
 \right).
 \]
Here, we have chosen $B_{i}(\mathbf{r}), \phi_{i}(\mathbf{r}) \in \mathbb{R}_{\geq 0}$, ($i=x,y,z$). 
In the following, we consider a fixed position in space and suppress the dependence of $B_{i}(\mathbf{r})$ and $\phi_{i}(\mathbf{r})$ on $\mathbf{r}$ to simplify notation.
In our experiment, we apply a homogeneous static magnetic field $\mathbf{B}_{0}$ along several directions in order to be able to reconstruct all components of the mw magnetic field.
For a given $\mathbf{B}_{0}$, we choose a new cartesian coordinate system $(x',y',z')$ with the $z'$-axis pointing along $\mathbf{B_0}$, which defines the quantization axis for the atomic states $\ket{F,m_{F}}$. 
In this new coordinate system, the mw magnetic field phasor is given by 
\[
\mathbf{B}\equiv\left(
\begin{array}{l}
	B_{x'}e^{-i\phi_{x'}} \\
	B_{y'}e^{-i\phi_{y'}} \\
	B_{z'}e^{-i\phi_{z'}} 
\end{array} 
\right). 
\]
The mw magnetic field couples to the magnetic moment of the electron spin of the atom. The coupling to the nuclear magnetic moment is neglected, because it is three orders of magnitude smaller than the electron magnetic moment.
The Rabi frequency on the hyperfine transition $\ket{1,m_1}\leftrightarrow\ket{2,m_2}$ is given by
\begin{equation} 
\Omega_{1,m_1}^{2,m_{2}}=\frac{2 \mu_{B}}{\hbar}\bra{2,m_{2}}\mathbf{B}\cdot{\mathbf{J}}\ket{1,m_1},
\end{equation}
with $\mathbf{J}=(J_{x'}, J_{y'}, J_{z'})$ the electron spin operator. Using $J_{\pm}=J_{x'}\pm i J_{y'}$ we can write
\begin{eqnarray}
\mathbf{B}\cdot \mathbf{J} &=& B_{x'}e^{-i\phi_{x'}}J_{x'}+B_{y'}e^{-i \phi_{y'}}J_{y'}+B_{z'}e^{-i\phi_{z'}}J_{z'} \\
                           &=& \tfrac{1}{2}\left( B_{x'}e^{-i\phi_{x'}}-i B_{y'}e^{-i \phi_{y'}} \right)J_{+}+\tfrac{1}{2}\left( B_{x'}e^{-i\phi_{x'}}+i B_{y'}e^{-i \phi_{y'}} \right)J_{-}+B_{z'}e^{-i \phi_{z'}}J_{z'}.
\end{eqnarray}
Evaluating the matrix elements for the three transitions connecting to $\ket{1,-1}$,\cite{TreutleinThesis08} we obtain the Rabi frequencies: 
\begin{eqnarray}
 \underline{\Omega_{-}} &\equiv& \Omega_{1,-1}^{2,-2} = \frac{2 \mu_{B}}{\hbar}\bra{2,-2}\frac{1}{2}\left( B_{x'}e^{-i \phi_{x'}}+i B_{y'}e^{-i \phi_{y'}}\right)J_{-}\ket{1,-1}=\underline{-e^{-i \phi_{-}} \cdot \sqrt{3}\cdot \frac{ \mu_{B}}{\hbar} B_{-}},\label{OmegaMinus}\\
\underline{\Omega_{\pi}} &\equiv& \Omega_{1,-1}^{2,-1} = \frac{2 \mu_{B}}{\hbar}\bra{2,-1}B_{z'}e^{-i \phi_{z'}}J_{z'}\ket{1,-1}=\underline{-e^{-i \phi_{\pi}} \cdot \sqrt{\frac{3}{4}}\cdot \frac{ \mu_{B}}{\hbar} B_{\pi}}, \label{OmegaPi} \\
\underline{\Omega_{+}} &\equiv& \Omega_{1,-1}^{2,0} = 
 \frac{2 \mu_{B}}{\hbar}\bra{2,0}\frac{1}{2}\left( B_{x'}e^{-i \phi_{x'}}-i B_{y'}e^{-i \phi_{y'}}\right)J_{+}\ket{1,-1}=\underline{e^{-i \phi_{+}} \cdot \sqrt{\frac{1}{2}}\cdot \frac{ \mu_{B}}{\hbar} B_{+}}, \label{OmegaPlus}
\end{eqnarray}
where we have used the definitions 
\begin{eqnarray}
B_{-}e^{-i \phi_{-}} &:=& \tfrac{1}{2}\left( B_{x'}e^{-i \phi_{x'}}+i B_{y'}e^{-i \phi_{y'}} \right), \label{BMinus}\\
B_{\pi}e^{-i \phi_{\pi}} &:=& B_{z'}e^{-i \phi_{z'}}, \label{BPi} \\
B_{+}e^{-i \phi_{+}}  &:=& \tfrac{1}{2}\left( B_{x'}e^{-i \phi_{x'}}-i B_{y'}e^{-i \phi_{y'}} \right), \label{BPlus}
\end{eqnarray} 
with  $B_{\gamma},\phi_{\gamma} \in \mathbb{R}_{\geq 0}$, ($\gamma= -, \pi,  +$). 
We note that $\Omega_{\pi}$ is proportional to the projection of $\mathbf{B}$ onto $\mathbf{B}_{0}$, while $\Omega_{+(-)}$ is proportional to the right (left) handed circular polarization component in the plane perpendicular to $\mathbf{B}_{0}$. 

In the experiment, we choose a sufficiently strong static field so that $\omega_{L} \gg \Omega_{\gamma}$. Furthermore, we choose the microwave frequency resonant with one of the transitions, $\omega=\omega_\gamma$. In this way, Rabi oscillations are induced only on the resonant transition in a given run of the experiment, which allows us to selectively image the individual microwave magnetic field components $B_\gamma$.

\subsection{Reconstruction of the microwave magnetic field \label{sec:reconstr}}
The amplitudes $B_{x}$, $B_{y}$, and $B_{z}$ of the cartesian components of $\mathbf{B}$ in laboratory coordinates can be easily determined by measuring $|\Omega_{\pi}|$ with the quantization axis $\mathbf{B}_{0}/B_{0}$ pointing along $x$, $y$, and $z$, respectively. The extraction of the field components from absorption images is described in the main text.	
In the following, the upper index indicates the direction of the quantization axis in laboratory coordinates, e.g.\ $\Omega_{-}^{y}$ ($B_{-}^{y}$) means $\Omega_{-}$ ($B_{-}$) for $\mathbf{B}_{0}$ pointing along the y-axis. 

To reconstruct the relative phases $\left(\phi_y-\phi_x\right)$ and $\left(\phi_z-\phi_x\right)$ between the cartesian components of $\mathbf{B}$, we also measure the amplitudes of the circularly polarized components $B_{+}^{x}$, $B_{-}^{x}$, $B_{+}^{y}$, $B_{-}^{y}$, $B_{+}^{z}$, and $B_{-}^{z}$. 
Having measured these components, we can reconstruct the relative phases according to the following recipe.
\subsubsection{$\mathbf{B}_{0}$ along $x$}
\label{sec:MathbfB0AlongX}
We choose a new coordinate system with $z'$ along $x$, resulting from the following coordinate transformation: \\
\begin{center}
  \begin{tabular}{|c|}
    \hline
    $\begin{array}{l}
  	  x' = -z \\
	    y' = y \\
	    z' = x 
    \end{array} $ \\
    \hline
  \end{tabular}
\end{center}
In this rotated coordinate system, the mw magnetic field phasor reads 
\[
\mathbf{B}\equiv\left(
\begin{array}{l}
	B_{x'}e^{-i\phi_{x'}} \\
	B_{y'}e^{-i\phi_{y'}} \\
	B_{z'}e^{-i\phi_{z'}} 
\end{array} 
\right) = \left(
\begin{array}{l}
	-B_{z}e^{-i\phi_{z}} \\
	B_{y}e^{-i\phi_{y}} \\
	B_{x}e^{-i\phi_{x}} 
\end{array} \right).
\]
From this we obtain:
\begin{center}
  \begin{tabular}{|clr|cl|}
    \hline
      $B_{x'}$&=&$B_{z}$  & $\phi_{x'}=$&$\phi_{z}+\pi$ \\
      $B_{y'}$&=&$B_{y}$  & $\phi_{y'}=$&$\phi_{y}$ \\
      $B_{z'}$&=&$B_{x}$ & $\phi_{z'}=$&$\phi_{x}$ \\    
    \hline
  \end{tabular}
\end{center}
Using Eqs.~(\ref{BMinus}) and (\ref{BPlus}), we calculate 
\begin{equation}
B_{+}^{2}-B_{-}^{2}=-B_{x'}B_{y'}\sin (\phi_{y'}-\phi_{x'}). \\
\end{equation}
By insertion of the coordinate transformation and using Eqs.~(\ref{OmegaMinus}) - (\ref{OmegaPlus}), we obtain 
\begin{equation}
  \sin\left( \phi_{z}-\phi_{y}  \right) = \frac{\hbar}{4 \mu_{B}^{2}B_{z}B_{y}}\left( \frac{4}{3} |\Omega_{ -}^{x}|^{2} - 8 |\Omega_{ +}^{x}|^{2}\right) = \frac{1}{\hbar B_{y} B_{z}}\left( (B_{-}^{x})^{2} - (B_{+}^{x})^{2}\right). \label{sinX}
\end{equation}
\subsubsection{$\mathbf{B}_{0}$ along $y$}
\label{sec:MathbfB0AlongY}
A similar calculation as before yields 
\begin{equation}
  \sin \left(  \phi_{x} - \phi_{z} \right) = \frac{\hbar}{4 \mu_{B}^{2}B_{x}B_{z}}\left( \frac{4}{3} |\Omega_{-}^{y}|-8 |\Omega_{ +}^{y}|^{2}\right)=\frac{1}{\hbar B_{x} B_{z}} \left((B_{-}^{y})^{2} - (B_{+}^{y})^{2}\right).  \label{sinZ}
\end{equation}
\subsubsection{$\mathbf{B}_{0}$ along $z$}
\label{sec:MathbfB0AlongZ}
In this case, we obtain
\begin{equation}
  \sin \left(  \phi_{y} - \phi_{x} \right) = \frac{\hbar}{4 \mu_{B}^{2}B_{x}B_{y}}\left( \frac{4}{3} |\Omega_{-}^{z}|-8 |\Omega_{ +}^{z}|^{2}  \right) = \frac{1}{\hbar B_{x} B_{y}} \left( (B_{-}^{z})^{2} - (B_{+}^{z})^{2}\right).  \label{sinY}
\end{equation}

All quantities on the right hand sides of Eqs.~(\ref{sinX}) - (\ref{sinY}) can be measured. 
From Eqs.~(\ref{sinX}) - (\ref{sinY}), the relative phases $(\phi_{y}-\phi_{x})$ and $(\phi_{z}-\phi_{x})$ can be determined. The solution is unique except for the very degenerate case where $\sin \left( \phi_{x}-\phi_{z}\right)=\sin \left( \phi_{y}-\phi_{x}\right)=\sin \left( \phi_{z}-\phi_{y}\right)=0$. In this case, there are 4 solutions which cannot be distinguished.

\subsection{Reconstruction of the absolute microwave phase}
While Sec.~\ref{sec:reconstr} describes the reconstruction of the relative phases $(\phi_{y}-\phi_{x})$ and $(\phi_{z}-\phi_{x})$ between the cartesian components of $\mathbf{B}$, it is also possible to reconstruct the spatial dependence of the global phase of $\mathbf{B}$.   
The procedure 
uses two mw pulses. 
During the whole sequence, $\mathbf{B}_{0}$ stays the same. In the following, we assume $\mathbf{B}_{0}$ is pointing along the x-axis, so that $\phi_{x}(\mathbf{r})$ is measured. The other two phases, $\phi_{y}(\mathbf{r})$ and $\phi_{z}(\mathbf{r})$, can then be determined from the already known relative phases.  

After releasing the atoms from the trap, they are prepared in an equal superposition of states $\ket{1,-1}$ and $\ket{2,-1}$ by application of a $\tfrac{\pi}{2}$-pulse at frequency $\omega = \omega_{\pi}$. This mw pulse is applied from a well-characterized source, so that it has negligible (or at least known) intensity gradients and negligible (or known) phase gradients across the atomic cloud. This can be achieved by using an external mw horn.\cite{Boehi09a}  The duration of the pulse is $dt_{\mathrm{mw1}}=\frac{\pi}{2 |\Omega_{\mathrm{\pi,1}}|}$, where $ \Omega_{\mathrm{\pi,1}}=\left|\Omega_{\mathrm{\pi,1}}\right|e^{i \phi_{\pi,1}}$ is the Rabi frequency of the pulse. The state after this preparation pulse is (in the rotating wave approximation) \cite{Scully97}
\begin{equation}
\ket{\psi_{1}}=\tfrac{1}{\sqrt{2}}\left( \ket{1,-1}+i e^{-i \phi_{\pi,1}} \ket{2,-1} \right). \label{eq:equalSuperpos} 
\end{equation}
Immediately after the end of this preparation pulse, the mw in the MMIC to be characterized is pulsed on at frequeny $\omega_{\mathrm{mw,2}}=\omega_{\pi}$ for a duration $dt_{\mathrm{mw,2}}$. The Rabi frequency and phase of this second mw pulse are denoted by $\Omega_{\pi,2}(\mathbf{r})$ and $\phi_{\pi,2}(\mathbf{r})$, respectively. After the second pulse, the state of an atom at position $\mathbf{r}$ is
\begin{eqnarray*}
\ket{\psi_{2}(\mathbf{r})}&=&\tfrac{1}{\sqrt{2}} \left[ \cos\left( \frac{\left|\Omega_{\pi,2}(\mathbf{r})\right|dt_{\mathrm{mw,2}}}{2}\right)- e^{i \phi_{\pi,2}-i\phi_{\pi,1}} \sin \left( \frac{\left|\Omega_{\pi,2}(\mathbf{r})\right|dt_{\mathrm{mw,2}}}{2} \right)   \right] \ket{1,-1} \\
&+& \tfrac{1}{\sqrt{2}}\left[ i e^{-i\phi_{\pi,1}}\cos\left( \frac{\left|\Omega_{\pi,2}(\mathbf{r})\right|dt_{\mathrm{mw,2}}}{2} \right) + i e^{-i \phi_{\pi,2}}\sin\left( \frac{\left|\Omega_{\pi,2}(\mathbf{r})\right|dt_{\mathrm{mw,2}}}{2} \right) \right]\ket{2,-1} .
\end{eqnarray*}
The probability $p_{2}(\mathbf{r})$ of finding an atom at position $\mathbf{r}$ in state $F=2$ is given by
\begin{equation}
p_{2}(\mathbf{r})=\frac{1}{2} + \frac{1}{2}\sin\left( \left| \Omega_{\pi,2}(\mathbf{r}) \right| dt_{\mathrm{mw,2}} \right)\cdot \cos\left( \phi_{\pi,1}-\phi_{\pi,2}(\mathbf{r}) \right)
\end{equation}
To calculate $\phi_{\pi,2}(\mathbf{r})$, the quantities $|\Omega_{\pi,2}(\mathbf{r})|$ and $\phi_{\pi,1}$ have to be known. $|\Omega_{\pi,2}(\mathbf{r})|$ can be measured as described in the main text. 
If $\mathbf{B}_{0}$ is pointing along the $x$-axis, then $\phi_{x}(\mathbf{r})\equiv \phi_{\pi,2}(\mathbf{r})$.

The calculation above assumes that there is zero delay between the end of the first preparation pulse and the second pulse in the MMIC.  
A similar calculation is also possible for non-zero delay between the two pulses.

\subsection{Sensitivity \& spatial resolution}
\label{sec:Sensitivity}
In this section we estimate the maximum sensitivity of our 
technique for our set of experimental parameters. 
As described in the main text, the mw magnetic field sensitivity is mainly determined by the interaction time $dt_{\mathrm{mw}}$ of the atoms with the mw pulse on the MMIC. A longer interaction time $dt_{\mathrm{mw}}$ leads to a higher mw magnetic field sensitivity, because then a weaker mw field can already drive a substantial fraction of a Rabi cycle. However, at the same time the effective spatial resolution  $s_{\mathrm{eff}} \equiv 2\sigma_{\mathrm{eff}}$ decreases as the image blurs due to the movement of the atoms during $dt_{\mathrm{mw}}$.  In the following, the symbol $\sigma$ always refers to an r.m.s.\ width.  $\sigma_{\mathrm{eff}}$ is determined by the average moving distance $\sigma_{\mathrm{mw}}$ of the atoms  during the mw pulse, by the optical resolution of the imaging system $s_{\mathrm{opt}} \equiv 2\sigma_{\mathrm{opt}}$, and by the movement of the atoms during the imaging laser pulse, where thermal motion ($\sigma_{\mathrm{tm}}$) and diffusive motion due to photon scattering ($\sigma_{\mathrm{ps}}$) contribute. In the following, we will calculate $\sigma_{\mathrm{mw}}$, $\sigma_{\mathrm{tm}}$, $\sigma_{\mathrm{ps}}$, and $\sigma_{\mathrm{eff}}$.

\subsubsection{Movement of atoms during the microwave pulse - $\sigma_{\mathrm{mw}}$}
\label{sec:sMathrmMw}
A free-falling atom in a cloud at temperature $T$ has a mean thermal velocity perpendicular to the line of sight of $v_{\mathrm{th}}=\sqrt{\frac{2 k_{B} T}{m}}$.\cite{Ketterle99} After releasing the atom from the trap and waiting for $dt_{\mathrm{ho}}$, the atom has furthermore acquired a velocity $v_g=g \cdot dt_{\mathrm{ho}}$ along the direction of gravity. During the interaction with the mw for a time $dt_{\mathrm{mw}}$, the atom moves ballistically by an average distance 
\begin{equation}
\sigma_{\mathrm{mw}} = g \cdot dt_{\mathrm{ho}}\cdot dt_{\mathrm{mw}}+\frac{1}{2}g\cdot dt_{\mathrm{mw}}^{2}+\sqrt{\frac{2k_{B} T}{m}}dt_{\mathrm{mw}}. \label{eq:dtmwmax}
\end{equation}
For $T=5~\mathrm{\mu K}$, $dt_{\mathrm{ho}}=0$, and $dt_{\mathrm{mw}}=80~\mathrm{\mu s}$ we obtain $\sigma_{\mathrm{mw}}=2.5~\mathrm{\mu m}$. 
For short times, the last term in the equation above dominates (as it is the case for our parameters). The displacement of the atoms during the mw pulse can then be approximated by a Gaussian function $f_{\mathrm{mw}}$ of r.m.s.\ width $\sigma_{\mathrm{mw}}$.


\subsubsection{Movement of atoms during the imaging pulse - $\sigma_{\mathrm{tm}}$}
During the imaging pulse of duration $dt_{\mathrm{im}}$, the atoms move ballistically by an average distance $\sigma_{\mathrm{tm}}$ given by
  \begin{equation}
    \sigma_{\mathrm{tm}} = g(dt_{\mathrm{ho}}+dt_{\mathrm{mw}})dt_{\mathrm{im}}+\frac{1}{2}g \cdot dt_{\mathrm{im}}^{2}+\sqrt{\frac{ 2k_{B} T}{m}} \cdot dt_{\mathrm{im}}
  \end{equation} 
  due to gravity and thermal motion.
  For 
  $dt_{\mathrm{im}}=40~\mathrm{\mu s}$ we get $\sigma_{\mathrm{tm}}=1.3~\mathrm{\mu m}$. 
  Again, for short times the atomic density distribution after the imaging pulse can be approximated by a Gaussian $f_{\mathrm{tm}}$ with r.m.s.\ width $\sigma_{\mathrm{tm}}$. 
  
\subsubsection{Diffusive movement of atoms due to photon scattering - $\sigma_{\mathrm{ps}}$}
\label{sec:RandomMovementOfAtomsDueToScatteringOfPhotonsSigmaMathrmIm}
During the imaging laser pulse of duration $dt_{\mathrm{im}}$, the atoms randomly scatter photons. The associated momentum recoils lead to a diffusive motion of the atoms, which leads at the end of the pulse to an average displacement perpendicular to the line of sight of \cite{Ketterle99}
\[
\sigma_{\mathrm{ps}}=\sqrt{\frac{2}{3}}\cdot\sqrt{\frac{N_{p}}{3}}\cdot v_{\mathrm{rec}} \cdot dt_{\mathrm{im}} ,
\]
where $v_{\mathrm{rec}}=\hbar k/m=5.9\,\mathrm{mm/s}$ is the atomic recoil velocity for $^{87}\mathrm{Rb}$, and $N_{p}$ the number of scattered photons, with $N_{p}=(\Gamma/2)\, dt_{\mathrm{im}} \,s/(1+s) $ and the natural linewidth $\Gamma=2\pi\times 6.1$~MHz. 
For our experimental parameters (saturation parameter $s=I/I_{\mathrm{sat}}=1$, $dt_{\mathrm{im}}=40~\mathrm{\mu s}$) we get $\sigma_{\mathrm{ps}}=2.2~\mathrm{\mu m}$. 

\subsubsection{Effective spatial resolution $s_\mathrm{eff}=2\sigma_{\mathrm{eff}}$}
\label{sec:EffectiveResolutionDMathrmEff}
The effective resolution can approximately be calculated by the convolution $f_{\mathrm{eff}}=f_{\mathrm{opt}}\ast \left[f_{\mathrm{ps}} \ast \left( f_{\mathrm{tm}} \ast f_{\mathrm{mw}}\right)\right]$, where $f_{\mathrm{opt}}$ approximates the point spread function of the imaging system by a Gaussian $f_{\mathrm{opt}}$ with $\sigma_{\mathrm{opt}}=2~\mathrm{\mu m}$.
As a result, we get $\sigma_{\mathrm{eff}}=4.1~\mathrm{\mu m}$. We take as the effective resolution for our parameters $s_{\mathrm{eff}}=2\sigma_{\mathrm{eff}}=8.2~\mathrm{\mu m}$.

\subsubsection{Imaging noise}
\label{sec:ImagingNoise}
The noise on the absorption images is important for the sensitivity of our technique. It determines the minimum number of atoms $N_{2,\mathrm{min}}$ that has to be transferred by the mw into the $F=2$ manifold during $dt_{\mathrm{mw}}$ in order for the mw field to be detectable.
We currently use an \textit{Andor iKon-M} camera with a quantum efficiency of 90\% for absorption imaging \cite{Note1}. The optical resolution of our imaging system is $s_{\mathrm{opt}}=4~\mathrm{\mu m}$, the imaging pulse duration $dt_{\mathrm{im}}=40~\mathrm{\mu s}$, and we are imaging at saturation intensity ($s=I/I_{\mathrm{sat}}=1$). With these parameters, we calculate an uncertainty in the number of atoms detected in an area
$A_\mathrm{eff}=\pi \sigma_{\mathrm{eff}}^{2}$ of 1.4 atoms r.m.s. We measure a value of $\sigma_{N,\mathrm{psn}}=2.0$~atoms. 
The difference can be explained by interference fringes on the image. This additional noise could certainly be decreased further.

Quantum projection noise due to the probabilistic nature of the measurement process is an additional contribution of noise on the images. The measurement process projects the atomic superposition state onto the $F=1$ and $F=2$ states, resulting in a number of atoms of $N_1$ and $N_2$ in the two states, respectively. Even if the total number of atoms $N=N_1+N_2$ is the same in each shot, $N_1$ and $N_2$ will show (anticorrelated) fluctuations. This projection noise has an r.m.s. amplitude of $\sigma_{N1}=\sigma_{N2}=\sqrt{N\cdot p_{2}\cdot (1-p_{2})}$, where $\sigma_{Ni}$ denotes noise in $N_i$ and $p_2 = N_2/N$. The total noise on $N_2$ is thus $\sigma_{N,\mathrm{tot}}=\sqrt{\sigma_{N,\mathrm{psn}}^{2}+\sigma_{N2}^{2}}$.
We find that in order to obtain a signal-to-noise-ratio $\mathrm{SNR} \equiv N_2/\sigma_{N,\mathrm{tot}}>1$, we have to have  $N_2 > N_{2,\mathrm{min}}=3$.

\subsubsection{Microwave field sensitivity}
\label{sec:MicrowaveFieldSensitivity}
In the experiment described in the main paper, we trap about $N=9\times 10^3$ atoms in the magnetic trap. The trapping frequencies are $\omega_{x}=2\pi \times 27~\mathrm{Hz}$ and $\omega_{y}\approx \omega_{z}=2\pi\times 680~\mathrm{Hz}$. We calculate the average atomic density in the trap to $n=2.2\times 10^{11}/\mathrm{cm}^{3}$.\cite{Ketterle99} The trapped cloud has a $1/e$-radius of $\rho = \sqrt{\frac{2k_{B}T}{m}}\frac{1}{\omega_y}=7.2~\mathrm{\mu m}$ along the $y$-axis, which is the direction of the imaging beam.
If we image the atoms with $dt_{\mathrm{ho}}=0$, we have about $N=170$ atoms inside a cylinder of radius $\sigma_{\mathrm{eff}}$ and height $2 \rho$.
The mw magnetic field which transfers on average $N_{2,\mathrm{min}}=3$ atoms to $F=2$ is obtained by requiring that 
\begin{equation}
N_2 = N \sin^{2}\left[ \frac{1}{2} |\Omega_\gamma| dt_{\mathrm{mw}} \right]\stackrel{!}{=}N_{2,\mathrm{min}}.\label{eq:OmMindestens}
\end{equation}    
For the three transitions $\omega=\omega_{ -}$, $\omega=\omega_{\pi}$, and $\omega=\omega_{ +}$, this is equivalent to
\begin{equation}
N_{F=2}=N \sin^{2}\left[ \frac{1}{2} \left( \frac{\mu_{B}}{\hbar}\sqrt{3}B_{-}\right) dt_{\mathrm{mw}} \right] \approx N \left[ \frac{1}{2} \left( \frac{\mu_{B}}{\hbar}\sqrt{3}B_{-}\right) dt_{\mathrm{mw}} \right]^{2}\stackrel{!}{=}3, \label{eq:BSig-Sens}
\end{equation}    
\begin{equation}
N_{F=2}=N \sin^{2}\left[ \frac{1}{2} \left( \frac{\mu_{B}}{\hbar}\sqrt{\frac{3}{4}}B_{\pi}\right) dt_{\mathrm{mw}} \right] \approx N \left[ \frac{1}{2} \left( \frac{\mu_{B}}{\hbar}\sqrt{\frac{3}{4}}B_{\pi}\right) dt_{\mathrm{mw}} \right]^{2}\stackrel{!}{=}3, \label{eq:BPiSens}
\end{equation}    
\begin{equation}
N_{F=2}=N \sin^{2}\left[ \frac{1}{2} \left( \frac{\mu_{B}}{\hbar}\sqrt{\frac{1}{2}}B_{+}\right) dt_{\mathrm{mw}} \right] \approx N \left[ \frac{1}{2} \left( \frac{\mu_{B}}{\hbar}\sqrt{\frac{1}{2}}B_{+}\right) dt_{\mathrm{mw}} \right]^{2}\stackrel{!}{=}3. \label{eq:BSig+Sens}
\end{equation}    
Solving the above equations with $dt_{\mathrm{mw}}= 80\,\mathrm{\mu s}$, we get $B_{-}=2.2\cdot 10^{-8}~\mathrm{T}$, $B_{\pi}=4.4\cdot 10^{-8}~\mathrm{T}$, and $B_{+}=5.4\cdot 10^{-8}~\mathrm{T}$. 


Note that the projection noise $\sigma_{N2}$ slowly increases with increasing $N_{2}$. Therefore, the absolute mw field resolution of our method decreases with increasing values of $\mathbf{B}$.

The consideration above relies on the assumption that we can achieve perfect resonance $\omega = \omega_{\gamma}$. 
Solving Eq.~(\ref{eq:OmMindestens}) for  $N_{2,\mathrm{min}}$, we obtain $|\Omega_\gamma|/2\pi = 0.53~\mathrm{kHz}$ for $N_{2}=3$. A change in $\omega_{\gamma}/2\pi$ of 0.53~kHz corresponds to a magnetic field instability of $2.5\times 10^{-8}$~T for $\Omega_{ -}$, $3.8\times 10^{-8}$~T for $\Omega_{\pi}$ and $7.6\times 10^{-8}$~T for $\Omega_{ +}$. Inside the magnetic shielding surrounding our experiment, we achieve a stability of $B_{0}$ of $2\times 10^{-8}$~T r.m.s., which could certainly be improved such that the effect can be neglected. 

The sensitivity of our field imaging technique can be increased by using colder or denser clouds. Suitable techniques to reduce the temperature further are adiabatic relaxation of the trap or further forced evaporative cooling.

\subsection{Measurement of the microwave field with trapped atoms}
\label{sec:InTrapMeasurementOfBMathrmMw}
There are further operation modes of our field imaging method. 
Instead of releasing the atoms from the trap, it is also possible to prepare a Bose-Einstein condensate (BEC) \cite{Ketterle99} in a trap (either magnetic or optical) or in an array of traps (e.g.\ an optical lattice) and scan its position spatially from shot to shot. The position can be scanned on a $\mathrm{\mu m}$ length scale  in all three dimensions. A typical stepsize is given by twice the Thomas-Fermi radius $r_{\mathrm{TF}}$ of the BEC.\cite{Ketterle99} If the BEC is held in a magnetic trap, transitions between $\ket{1,-1}$ and $\ket{2,m_{2}}$ are detected via loss of atoms, because $\ket{2,m_{2}}$ with $m_2 = -2, -1, 0$ are magnetically non-trappable states. If the BEC(s) are held in an optical trap or optical lattice, then transitions to the $F=2$ manifold can be detected via state-selective imaging. 

\subsection{Tunability of transition frequencies $\omega_{ -}$, $\omega_{\pi}$ and $\omega_{ +}$}

\subsubsection{$^{87}$Rubidium atoms}
\label{sec:87MathrmRb}
The transition frequencies $\omega_{ -}$, $\omega_{\pi}$, and $\omega_{ +}$ between the initial state $\ket{1,-1}$ and the target states $\ket{2,-2}$, $\ket{2,-1}$, and $\ket{2,0}$, respectively, can be adjusted by changing the static magnetic field $B_{0}$. For small magnetic fields (i.e.\ $\omega_{L}\ll 2\pi \times 6.8~\mathrm{GHz}$) the transition frequencies are approximately given by  $\omega_{ -}=\omega_{0}-3 \omega_{L}$, $\omega_{\pi}=\omega_{0}-2 \omega_{L}$ and $\omega_{ +}=\omega_{0}- \omega_{L}$, where $\omega_{0}=2\pi\times 6834.682610~\mathrm{MHz}$ and $\omega_{L}=\mu_{B}B_{0}/2 \hbar$.

\begin{figure}[htbp]
	\centering
		\includegraphics{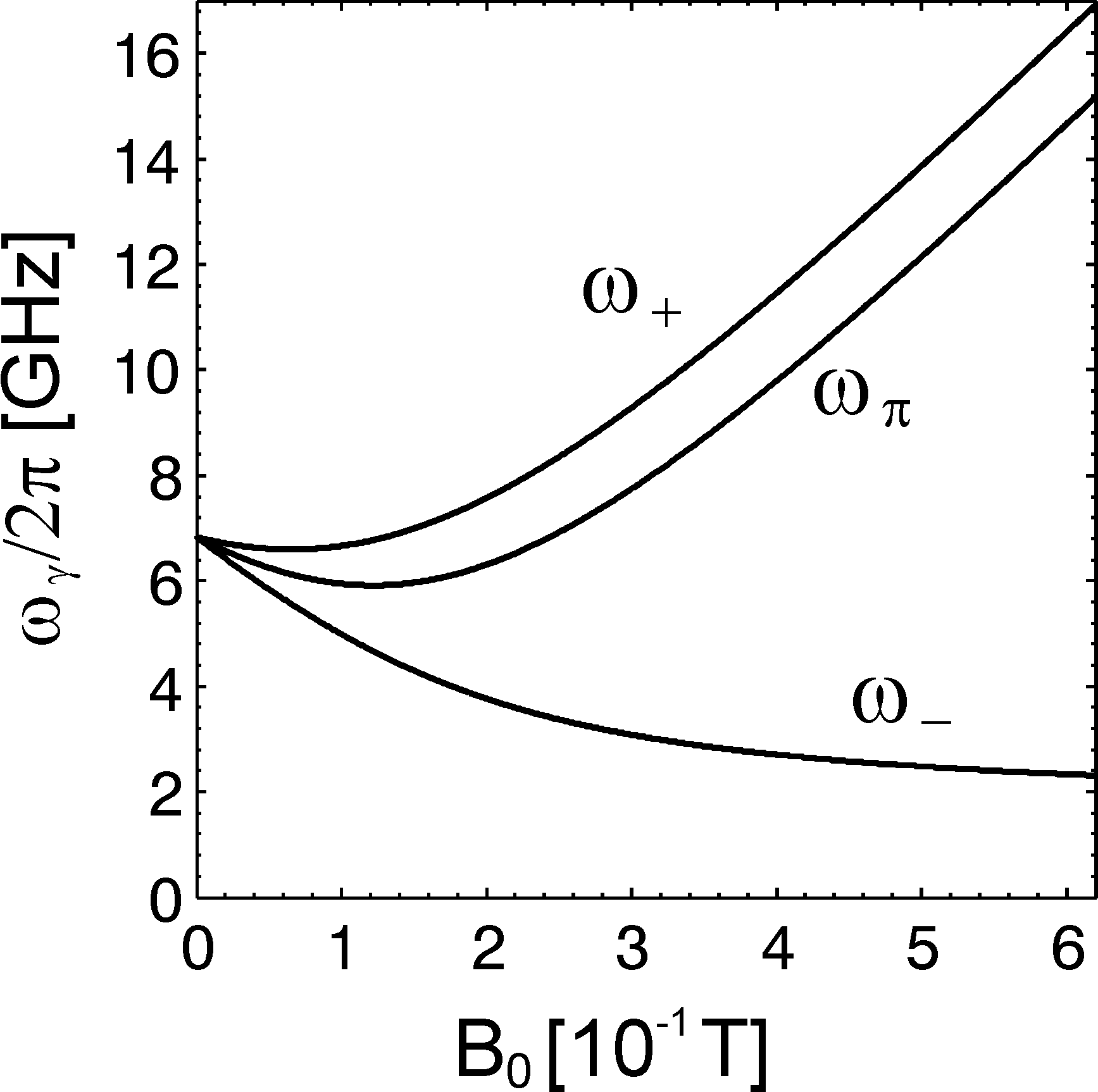}
	\caption{Transition frequencies $\omega_{ -}$, $\omega_{\pi}$ and $\omega_{ +}$ for $^{87}\mathrm{Rb}$ as a function of $B_{0}$. The individual transition frequencies can be tuned by up to 10~GHz with magnetic fields of up to $0.5~\mathrm{T}$.}
	\label{fig:BreitRabi}
\end{figure}

For larger values of $B_{0}$ a more accurate calculation is required. The $5^{2}S_{1/2}$  ground state hyperfine structure of $^{87}\mathrm{Rb}$ is described by the Hamiltonian \cite{TreutleinThesis08}
\begin{equation}
H = A_{\mathrm{hfs}} \mathbf{I}\cdot \mathbf{J}+\mu_{B}B_{0}(g_{J}J_{z}+g_{I}I_{z}), \label{eq:BreitRabi0}
\end{equation}
with $\mathbf{I}$ the nuclear spin operator and $\mathbf{J}$ the operator for the total electronic spin, $A_{\mathrm{hfs}}=\hbar \omega_0/2$, $g_{I}=-0.000995141$, and $g_{J}=2.002331$.\cite{Steck09} Even though $\frac{g_{I}}{g_{J}} \approx 5\cdot 10^{-4}$, the coupling of $\mathbf{I}$ to the magnetic field becomes important for large values of $B_{0}$. An analytical formula exists for the ground state manifold of a $D$ transition (as it is the case for our states of interest), the Breit-Rabi formula \cite{Steck09}:
\begin{equation}
E_{F,m_{F}}=-\frac{ A_{\mathrm{hfs}}}{(2I+1)}+g_{I}\mu_{B}m_{F}B\pm A_{\mathrm{hfs}}\left(1+\frac{4m_{F}\beta}{2I+1}+\beta^{2}\right)^{1/2},\label{eq:BreitRabi1}
\end{equation} 
where 
$\beta=\frac{(g_{J}-g_{I})\mu_{B}B_{0}}{2A_{\mathrm{hfs}}}$, $m_{F}=m_{I}\pm m_{J}$ (the $\pm$ sign is the same as in eq. \ref{eq:BreitRabi1}).
The resulting transition frequencies $\omega_{ -}$, $\omega_{\pi}$, and $\omega_{ +}$ as a function of $B_{0}$ are illustrated in Fig. \ref{fig:BreitRabi}.
For $^{87}\mathrm{Rb}$, the transition frequencies $\omega_{\gamma}/2\pi$ can be tuned over a range of more than 10~GHz using technically feasible magnetic fields. Note that for $B_{0}>0.1$~T we start entering the Paschen-Back regime, where the matrix elements of Eqs.~(\ref{OmegaMinus}) - (\ref{OmegaPlus}) change and the theory has to be modified. 

\subsubsection{Other atomic species}
\label{sec:OtherAtoms}
By using atomic species other than $^{87}\mathrm{Rb}$, different frequency ranges become accessible, e.g. $ 9.2~\mathrm{GHz}$ for Cs or $ 1.7~\mathrm{GHz}$ for Na.

\subsection{Ramsey interferometry and off-resonant probing}
\label{sec:EInterferometryAndOffResonantProbing}
Instead of having $\omega$ resonant with a hyperfine transition frequency, it is also possible to probe an off-resonant mw or light field (or anything else that causes a differential energy shift between the involved hyperfine levels) using a scheme based on Ramsey interferometry.\cite{Ramsey56} In Ramsey interferometry, the interaction with the mw field of duration $T_{\mathrm{int}}$ is enclosed by two $\frac{\pi}{2}$-pulses. The first pulse prepares the atoms in an equal superposition of two hyperfine states such as
\begin{equation}
\ket{\psi_{1}}(t=0)=\tfrac{1}{\sqrt{2}}\left( \ket{1,-1}+i e^{-i\phi_{\mathrm{mw}}} \ket{2,m_{2}} \right).
\end{equation}
During time $T_{\mathrm{int}}$ both states accumulate a differential phase shift $\Delta \phi=\int_{0}^{T_{\mathrm{int}}}\frac{E_{\mathrm{diff}}}{\hbar} dt$. $E_{\mathrm{diff}}$ is the differential potential between the two hyperfine levels involved and is in general state-dependent. The state after this interaction is
\begin{equation}
\ket{\psi_{1}}(t=T_{\mathrm{int}})=\tfrac{1}{\sqrt{2}}\left( \ket{1,-1}+i e^{-i\Delta 	\phi -i \phi_{\mathrm{mw}}} \ket{2,m_{2}} \right). 
\end{equation}
After applying the second $\frac{\pi}{2}$-pulse, the state is 
\begin{equation}
\ket{\psi_{2}}=\tfrac{1}{2}\left[\left( 1-e^{-i \Delta \phi} \right)\ket{1,-1}+ ie^{-i \phi_{\mathrm{mw}}}\left(  e^{-i \Delta \phi}+1 \right)\ket{2,m_{2}}\right],
\end{equation}
where $\phi_{\mathrm{mw}}$ is the phase of the resonant mw in a frame rotating at the atomic transition frequency. 
The probabilities to detect an atom in state $\ket{1,-1}$ and $\ket{2,m_{2}}$ are then given by 
\begin{eqnarray}
p_{1}(\mathbf{r})&=&\tfrac{1}{2}\left( 1-\cos(\Delta \phi) \right),  \\
p_{2}(\mathbf{r})=1-p_{1}(\mathbf{r})&=&\tfrac{1}{2}\left( 1+\cos(\Delta \phi) \right).  
\end{eqnarray}
By measuring the relative populations $p_{1}(\mathbf{r})$ and $p_{2}(\mathbf{r})$ after the second $\frac{\pi}{2}$ pulse, it is possible to determine the value of $\Delta \phi$ and thereby $E_{\mathrm{diff}}$. Details on the effect of an off-resonant mw field are discussed in \cite{Treutlein06b}.


\begin{thebibliography}{10}
\expandafter\ifx\csname url\endcsname\relax
  \def\url#1{\texttt{#1}}\fi
\expandafter\ifx\csname urlprefix\endcsname\relax\def\urlprefix{URL }\fi
\providecommand{\bibinfo}[2]{#2}
\providecommand{\eprint}[2][]{\url{#2}}


\bibitem{Robertson01}
\bibinfo{author}{I.D. Robertson}, and \bibinfo{author}{S. Lucyszyn}
\newblock \emph{\bibinfo{title}{RFIC and MMIC design and technology}}
  (\bibinfo{publisher}{The Institution of Electrical Engineers}, \bibinfo{address}{London},
  \bibinfo{year}{2001}), \bibinfo{edition}{1st} edn.  

\bibitem{DiCarlo09}
\bibinfo{author}{L. DiCarlo}, \bibinfo{author}{J.M. Chow}, \bibinfo{author}{J.M. Gambetta}, \bibinfo{author}{L.S. Bishop}, \bibinfo{author}{B.R. Johnson}, \bibinfo{author}{D.I. Schuster}, \bibinfo{author}{J. Majer}, \bibinfo{author}{A. Blais}, \bibinfo{author}{L. Frunzio}, \bibinfo{author}{S.M. Girvin}, and \bibinfo{author}{R.J. Schoelkopf}
\newblock \emph{\bibinfo{journal}{Nature}} \textbf{\bibinfo{volume}{460}},
\bibinfo{pages}{240-244} (\bibinfo{year}{2009}).



\bibitem{Boehi09a}
\bibinfo{author}{P. Böhi}, \bibinfo{author}{M.F. Riedel}, \bibinfo{author}{J. Hoffrogge}, \bibinfo{author}{J. Reichel}, \bibinfo{author}{T.W. Hänsch}, and \bibinfo{author}{P. Treutlein}
\newblock \emph{\bibinfo{journal}{Nat. Phys.}} \textbf{\bibinfo{volume}{5}},
  \bibinfo{pages}{592-597} (\bibinfo{year}{2009}).
  
  
\bibitem{Boehm94}
\bibinfo{author}{C. Böhm}, \bibinfo{author}{C. Roths}, and \bibinfo{author}{E. Kubalek} 
\newblock \emph{\bibinfo{journal}{Microwave Symposium Digest IEEE MTT-S International, San Diego}},
 (\bibinfo{year}{1994}).
    
    
\bibitem{imagingpapers}
\bibinfo{author}{S.K. Dutta}, \bibinfo{author}{C.P. Vlahacos}, \bibinfo{author}{D.E. Steinhauer}, \bibinfo{author}{A.S. Thanawala}, \bibinfo{author}{B.J. Feenstra}, \bibinfo{author}{F.C. Wellstood}, \bibinfo{author}{S.M. Anlage}, and \bibinfo{author}{H.S. Newman}
\newblock \emph{\bibinfo{journal}{Appl. Phys. Lett.}} \textbf{\bibinfo{volume}{74}}
\bibinfo{pages}{156-158} (\bibinfo{year}{1999}).  
\bibinfo{author}{C. Böhm}, \bibinfo{author}{F. Saurenbach}, \bibinfo{author}{P. Taschner}, \bibinfo{author}{C. Roths}, and \bibinfo{author}{E. Kubalek}
\newblock \emph{\bibinfo{journal}{J. Phys. D.: Appl. Phys.}} \textbf{\bibinfo{volume}{26}}
\bibinfo{pages}{1801-1805} (\bibinfo{year}{1993}).    
\bibinfo{author}{Y. Gao}, and \bibinfo{author}{I. Wolff},
\newblock \emph{\bibinfo{journal}{IEEE TMTT}} \textbf{\bibinfo{volume}{46}}
\bibinfo{pages}{907-913} (\bibinfo{year}{1998}).     
\bibinfo{author}{Y. Gao}, and \bibinfo{author}{I. Wolff},
\newblock \emph{\bibinfo{journal}{Microwave Symposium Digest, IEEE MTT-S International}} \textbf{\bibinfo{volume}{3}}
\bibinfo{pages}{1159-1162} (\bibinfo{year}{1995}).    
\bibinfo{author}{G. David}, \bibinfo{author}{P. Bussek}, \bibinfo{author}{U. Auer}, \bibinfo{author}{F.J. Tegude}, and \bibinfo{author}{D. Jäger}
\newblock \emph{\bibinfo{journal}{Electronic Letters}} \textbf{\bibinfo{volume}{31}}
\bibinfo{pages}{2188-2189} (\bibinfo{year}{1995}).      
\bibinfo{author}{T. Dubois}, \bibinfo{author}{S. Jarrix}, \bibinfo{author}{A. Penarier}, and \bibinfo{author}{P. Nouvel}
\newblock \emph{\bibinfo{journal}{IEEE Trans on Instrumentation and Measurement}} \textbf{\bibinfo{volume}{57}}
\bibinfo{pages}{2398-2404} (\bibinfo{year}{2008}).   
\bibinfo{author}{T.P. Budka}, \bibinfo{author}{S.D. Waclawik}, and \bibinfo{author}{G.M. Rebeiz}
\newblock \emph{\bibinfo{journal}{IEEE TMTT}} \textbf{\bibinfo{volume}{44}}
\bibinfo{pages}{2174-2184} (\bibinfo{year}{1996}).     
\bibinfo{author}{R.C. Black}, \bibinfo{author}{F.C. Wellstood}, \bibinfo{author}{E. Dantsker}, \bibinfo{author}{A.H. Miklich}, \bibinfo{author}{D.T. Nemeth}, \bibinfo{author}{D. Koelle}, \bibinfo{author}{F. Ludwig}, and \bibinfo{author}{J. Clarke}
\newblock \emph{\bibinfo{journal}{Appl. Phys. Lett.}} \textbf{\bibinfo{volume}{66}}
\bibinfo{pages}{1267-1269} (\bibinfo{year}{1995}).       
    
%
%
%
%
%
%
%
%
  
\bibitem{NatureInsight02}
\bibinfo{author}{S. Chu} 
\newblock \emph{\bibinfo{journal}{Nature}} \textbf{\bibinfo{volume}{416}}
\bibinfo{pages}{205-246} (\bibinfo{year}{2002}).   
 
\bibitem{Gentile89}
\bibinfo{author}{T.R. Gentile}, \bibinfo{author}{B.J. Hughey}, \bibinfo{author}{D. Kleppner}, and \bibinfo{author}{T.W. Ducas},
\newblock \emph{\bibinfo{journal}{Phys. Rev. A}} \textbf{\bibinfo{volume}{40}}
\bibinfo{pages}{5103-5115} (\bibinfo{year}{1989}).    
  
  
\bibitem{Matthews98}
\bibinfo{author}{M.R. Matthews}, \bibinfo{author}{D.S. Hall}, \bibinfo{author}{D.S. Jin}, \bibinfo{author}{J.R. Ensher}, \bibinfo{author}{C.E. Wieman}, \bibinfo{author}{E.A. Cornell}, \bibinfo{author}{F. Dalfovo}, \bibinfo{author}{C. Minniti}, and \bibinfo{author}{S. Stringari}
\newblock \emph{\bibinfo{journal}{Phys. Rev. Lett.}} \textbf{\bibinfo{volume}{81}}
\bibinfo{pages}{243-247} (\bibinfo{year}{1998}).       
    
 
  
  
  \bibitem{Ramsey56}
\bibinfo{author}{N.F. Ramsey}
\newblock 
\emph{\bibinfo{title}{Molecular Beams}}
  (\bibinfo{publisher}{Clarendon Press}, \bibinfo{address}{Oxford},
  \bibinfo{year}{1956}).
    
  
\bibitem{Buecker09}
\bibinfo{author}{R. Bücker}, \bibinfo{author}{A. Perrin}, \bibinfo{author}{S. Manz}, \bibinfo{author}{T. Betz}, \bibinfo{author}{C. Koller}, \bibinfo{author}{T. Plisson}, \bibinfo{author}{J. Rottmann}, \bibinfo{author}{T. Schumm}, and \bibinfo{author}{J. Schmiedmayer}
\newblock \emph{\bibinfo{journal}{New J. Phys.}} \textbf{\bibinfo{volume}{11}}
\bibinfo{pages}{103039} (\bibinfo{year}{2009}).    
    
  
  


\bibitem{Farkas09}
\bibinfo{author}{D.M. Farkas}, \bibinfo{author}{K.M. Hudek}, \bibinfo{author}{E.A. Salim}, \bibinfo{author}{S.R. Segal}, \bibinfo{author}{M.B. Squires}, and \bibinfo{author}{D.Z. Anderson}
\newblock \emph{\bibinfo{journal}{Appl. Phys. Lett.}} \textbf{\bibinfo{volume}{96}}
\bibinfo{pages}{093102} (\bibinfo{year}{2009}).     

  

\end{thebibliography}

\begin{thebibliography}{10}
\expandafter\ifx\csname url\endcsname\relax
  \def\url#1{\texttt{#1}}\fi
\expandafter\ifx\csname urlprefix\endcsname\relax\def\urlprefix{URL }\fi
\providecommand{\bibinfo}[2]{#2}
\providecommand{\eprint}[2][]{\url{#2}}

\bibitem{Gentile89}
\bibinfo{author}{T.R. Gentile}, \bibinfo{author}{B.J. Hughey}, \bibinfo{author}{D. Kleppner}, and \bibinfo{author}{T.W. Ducas},
\newblock \emph{\bibinfo{journal}{Phys. Rev. A}} \textbf{\bibinfo{volume}{40}}
\bibinfo{pages}{5103-5115} (\bibinfo{year}{1989}).  

\bibitem{TreutleinThesis08}
\bibinfo{author}{P. Treutlein}
\newblock Ph.D. thesis, \bibinfo{school}{Ludwig-Maximilians-Universit{\"a}t
  M{\"u}nchen and Max-Planck-Institut f{\"u}r Quantenoptik}
  (\bibinfo{year}{2008}).
\newblock \bibinfo{note}{Published as {MPQ} report 321}.


\bibitem{Boehi09a}
\bibinfo{author}{P. Böhi}, \bibinfo{author}{M.F. Riedel}, \bibinfo{author}{J. Hoffrogge}, \bibinfo{author}{J. Reichel}, \bibinfo{author}{T.W. Hänsch}, and \bibinfo{author}{P. Treutlein}
\newblock \emph{\bibinfo{journal}{Nat. Phys.}} \textbf{\bibinfo{volume}{5}},
  \bibinfo{pages}{592-597} (\bibinfo{year}{2009}).

\bibitem{Scully97}
\bibinfo{author}{M.O. Scully}, and \bibinfo{author}{M.S. Zubairy}
\newblock \emph{\bibinfo{title}{Quantum Optics}}
  (\bibinfo{publisher}{Cambridge University Press}, \bibinfo{address}{Cambridge, U.K.},
  \bibinfo{year}{1997}).
  
  \bibitem{Ketterle99}
\bibinfo{author}{W. Ketterle}, \bibinfo{author}{D.S. Durfee}, and  \bibinfo{author}{D.M. Stamper-Kurn}  
\newblock \bibinfo{title}{Making, probing and understanding {{B}ose-{E}instein} condensates}.
(\bibinfo{publisher}{IOS Press}, \bibinfo{address}{Amsterdam}, \bibinfo{year}{1999}).
  
  \bibitem{Note1}
  The images in the main paper were taken with another camera with lower quantum efficiency. For the estimate of the achievable resolution presented here, we use the parameters of the state-of-the-art camera that we currently employ in the experiment.
  
  
  \bibitem{Steck09}
\bibinfo{author}{D.A. Steck}
\newblock \emph{\bibinfo{journal}{http://steck.us/alkalidata/}}, version 2.1.2 (\bibinfo{year}{2009}).
  
  \bibitem{Ramsey56}
\bibinfo{author}{N.F. Ramsey}
\newblock \emph{\bibinfo{title}{Molecular Beams}}
  (\bibinfo{publisher}{Clarendon Press}, \bibinfo{address}{Oxford},
  \bibinfo{year}{1956}).
  
  
  \bibitem{Treutlein06b}
\bibinfo{author}{P. Treutlein}, \bibinfo{author}{T.W. Hänsch}, \bibinfo{author}{J. Reichel}, \bibinfo{author}{A. Negretti}, \bibinfo{author}{M.A. Cirone}, and \bibinfo{author}{T. Calarco},   
\newblock \emph{\bibinfo{journal}{Phys. Rev. A}} \textbf{\bibinfo{volume}{74}},
  \bibinfo{pages}{022312} (\bibinfo{year}{2006}).



\end{thebibliography}
\end{document}